\documentclass[english,aps,prl,twocolumn]{revtex4-1}
\usepackage[T1]{fontenc}
\usepackage[latin9]{inputenc}
\usepackage{amssymb}
\usepackage{graphicx}
\usepackage{babel}
\begin{document}

\title{Efficient algorithm for boson sampling with partially distinguishable
photons}

\author{J.J. Renema, A. Menssen, W.R. Clements, G. Triginer, W.S. Kolthammer
and I.A. Walmsley}

\affiliation{Clarendon Labs, Department of Physics, Oxford University, Parks Road
OX1 3PU Oxford}
\email{jelmer.renema@physics.ox.ac.uk}

\selectlanguage{english}%
\begin{abstract}
We demonstrate how boson sampling with photons of partial distinguishability
can be expressed in terms of interference of fewer photons. We use
this observation to propose a classical algorithm to simulate the
output of a boson sampler fed with photons of partial distinguishability.
We find conditions for which this algorithm is efficient, which gives
a lower limit on the required indistinguishability to demonstrate
a quantum advantage. Under these conditions, adding more photons only
polynomially increases the computational cost to simulate a boson
sampling experiment.
\end{abstract}
\maketitle
Boson sampling \cite{Aaronson2011} provides a promising route towards
demonstrating a quantum advantage, i.e. a computation by a quantum
system that exceeds what is possible with a classical one. In boson
sampling, the task is to provide a sample from the output of a linear
transformation of optical modes, some of which are fed with single
photons. For a sufficient number of photons and modes, a suitable
quantum machine directly implementing this problem will outperform
a realistic classical computer simulating the experiment. This result
has spurred a range of experimental efforts \cite{Spring2012,Broome2012,Crespi2013,Tillmann2013,Bentivegna2015,Carolan2015,Wang2017}. 

A crucial challenge for computational problems based on boson sampling
is to accommodate imperfections that arise in real-world devices.
An essential aspect of the original proposal \cite{Aaronson2011}
was to show that for small deviations from the ideal machine, the
achieved sampling problem retains computational hardness. However,
descriptions of experiments that incorporate realistic models of photon
distinguishability \cite{Shchesnovich2015a} or beam-splitter deviations
\cite{Arkhipov15} are unable to meet this error requirement. There
is therefore a need to devise photon sampling problems with improved
error tolerance. 

There exist two approaches to demarcating the line between viable
and non-viable extensions of boson sampling. In the top-down approach,
the original complexity proof is extended 'downwards', showing that
the resulting output distribution for systems of increasing imperfection
is still hard to sample from. This approach has had some success:
for example, it has been shown that the issue of unreliable single
photon generation can be overcome, and that the hardness of the resulting
'scattershot' boson sampling problem is equivalent to that of the
original boson sampling problem \cite{Aaronsonweb,Lund2014}. 

The alternate bottom-up approach is to construct new efficient classical
algorithms for boson samplers with particular imperfections. This
approach leads to constructive proofs that rule out a computational
advantage, thereby showing performance limits that must be exceeded
by quantum machines. For example, Rahimi-Keshari \emph{et al} \cite{RahimiKeshari2016}
used generalizations of the Wigner function as a way to construct
a classical algorithm which can efficiently simulate certain lossy
boson samplers. 

\begin{figure}
\includegraphics[width=8cm]{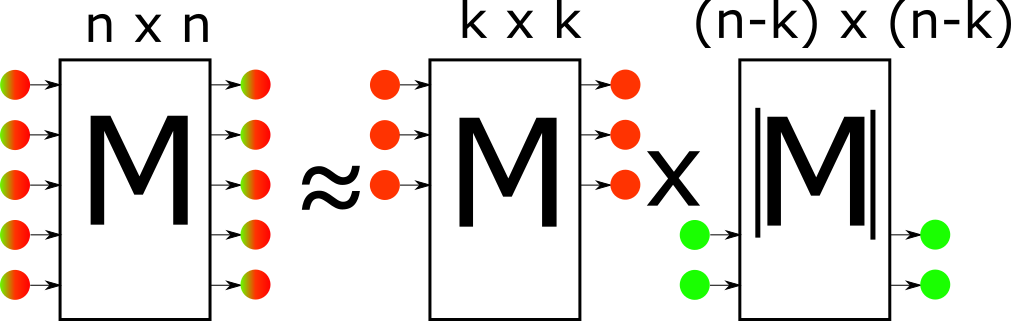}

\caption{A pictorial representation of our result. We show that boson sampling
with $n$ photons of partial distinguishability (represented by the
mixed red-green balls) can be approximated as computing the outcome
of a series of smaller permanents of size $k$, combined with probabilistic
transmission of the remaining $n-k$ photons. The value of $k$ at
which this approximation works is set by the value of the distinguishability.}
\end{figure}

In this work, we consider which-way information of the interfering
photons as an imperfection that compromises the hardness of a boson
sampler. We show a classical algorithm that efficiently approximates
detection probabilities in the limit of many photons, given the photons
have some partial distinguishability. We then use this algorithm to
consider problems of finite size and estimate a level of indistinguishability
that must be surpassed to demonstrate of quantum advantage. The basis
of our algorithm is, schematically depicted in Figure 1, that for
partially distinguishable photons, the probability of a given outcome
can be approximated by terms that involve fewer interfering photons,
where the remaining ones do not interfere at all. For a given error
tolerance and indistinguishability, we determine a number of photons
above which this approach succeeds while requiring only a polynomial
increase in the computational steps as the number of photons increases
further. We use this result to estimate a lower bound on the photon
quality required to demonstrate a quantum advantage: for 50 photons
and an error threshold of 10\%, the degree of indistinguishability
of the interfering photons must be higher than 94.7\%.

Multiphoton interference at partial distinguishability has been studied
extensively \cite{Shchesnovich2014,Shchesnovich2015a,Shchesnovich2015b,Tichypartial,Rohde2015,Tamma2015,Tamma2015PRL,Tamma2017,Tillmann2015}.
For boson sampling with fully indistinguishable photons, the probability
of a particular detection outcome is given by $P=\mathrm{|Perm}(M)|^{2},$
where $M$ is a submatrix of the unitary $U$ , where the rows and
columns of $M$ are chosen to correspond to the input and output modes
of interest, respectively. In this work, we will use the formalism
of Tichy \cite{Tichypartial}, where the probability of a particular
detection outcome (i.e. photons emerging at particular outputs) is:
\begin{equation}
P=\sum_{\sigma\in\Sigma}\left(\prod_{j}S_{\sigma_{j}j}\right)\mathrm{Perm}(M*M_{1,\mathbb{\sigma}}^{*}),
\end{equation}
where $M$ a submatrix constructed in the same way as for fully indistinguishable
boson sampling, $*$ denotes the elementwise product, and $M^{*}$
is the elementwise complex conjugation. The notation $M_{1,\sigma}$
indicates that the rows of $M$ are unpermuted, and that the columns
are permuted according to $\sigma$, and we will use this notation
convention throughout. The matrix of mutual distinguishabilities $S$
is given by $S_{ij}=\langle\Psi_{i}|\Psi_{j}\rangle$, where $\Psi_{i}$
is the $i$-th single-photon wavefunction. The set of permutations
of size $n$ is denoted $\Sigma$. 

There are two extreme cases to note. First, if $S_{ij}=1,$ eq. 1
reduces to the standard expression for boson sampling with fully indistinguishable
particles. For $S_{ij}=\delta_{ij},$ eq. 1 reduces to $P=\mathrm{Perm}(|M|^{2})$,
which is the expression for boson sampling with distinguishable photons.
In this latter case, multiphoton interference is absent and the total
probability is expressed in terms of single-photon transmission probabilities
instead of transmission amplitudes. Since this matrix contains only
positive elements, it can be evaluated to within a multiplicative
error in polynomial time \cite{Jerrum2001}. In our work, we will
interpolate between these cases, and parameterize the mutual distinguishabilities
by a single parameter $x,$ with $S_{ij}(x)=x+(1-x)\delta_{ij}$.
We will argue at the end of our work that our results apply to more
general forms of $S.$ 

The observation that underlies our work is that the degree of quantum
interference in each term in eq. 1 is determined by the number of
fixed points (invariant elements) in the corresponding permutation.
Each fixed point in $\sigma$ causes the corresponding row from $M$
to enter into the permanent as the modulus squared, meaning that in
that term, the corresponding photon does not exhibit interference.
Therefore, the size of the matrix of complex elements which must be
computed to evaluate each term is set by the number of non-invariant
elements in the permutation \cite{Rohde2015,Tichypartial}. Furthermore,
terms with many fixed points have a larger weight in the sum: for
each element in the permutation which is not a fixed point, the product
in eq. 1 will pick up a factor $x<1$ from the off-diagonal elements
of $S$ \cite{Shchesnovich2015a,Shchesnovich2015b}. One can therefore
construct a series of succesively more accurate approximations by
grouping the terms in eq. 1 by number of fixed points, and then truncating
the sum at some value, which we designate $k$. The resulting approximation
$P_{k}$ is given by: 
\begin{equation}
P_{k}=\sum_{j=0}^{k}\sum_{\sigma^{j}}x^{j}\mathrm{Perm}(M*M_{1,\sigma}^{*}),
\end{equation}
where we have introduced the notation $\sigma^{j}$ to denote those
permutations which have $n-j$ elements as fixed points. To simplify
eq. 2, we note that this permanent, $n-j$ columns will be left unpermuted,
and will therefore end up as the modulus squared of the elements.
Expanding eq. 2 in all the permuted columns and combining terms, we
can separate the permanents of permuted and unpermuted rows as:

\begin{equation}
P_{k}=\sum_{j=0}^{k}\sum_{\sigma^{j}}x^{j}\sum_{\rho}\mathrm{Perm}(M_{\rho,1}*M_{\rho,\sigma_{p}}^{*})\mathrm{Perm}(|M_{\bar{\rho},\sigma_{u}}|^{2}),
\end{equation}
where $\rho$ denotes the $\left(\begin{array}{c}
n\\
j
\end{array}\right)$ possible combinations of $j$ columns from the matrix $M,$ $\bar{\rho}$
denotes the complemenary rows, and $\sigma_{p}$ and $\sigma_{u}$
denote the permuted and unpermuted elements of $\sigma,$ respectively. 

We note that the lower the value of $j,$ the easier the terms are
to compute, since the second of these two permanents can be efficiently
evaluated, and the first permanent is of size $j.$ The term with
$j=0$ represents the case where the photons are treated as fully
distinguishable particles. The next term $(j=2)$ represents the first-order
correction, where interference between each pair of photons is considered,
and similarly for higher values of $j.$ The sequence of $P_{k}$
is therefore ordered by computational cost as well as by degree of
quantum interference.

The rest of this paper is dedicated to investigating the properties
of these succesive approximations. We start by investigating these
approximations numerically, which will lead to some conjectures regarding
the scaling behaviour of these approximations with $n$ and $k$,
which we will then confirm through more rigorous analysis.

\begin{figure}
\includegraphics[width=8.8cm]{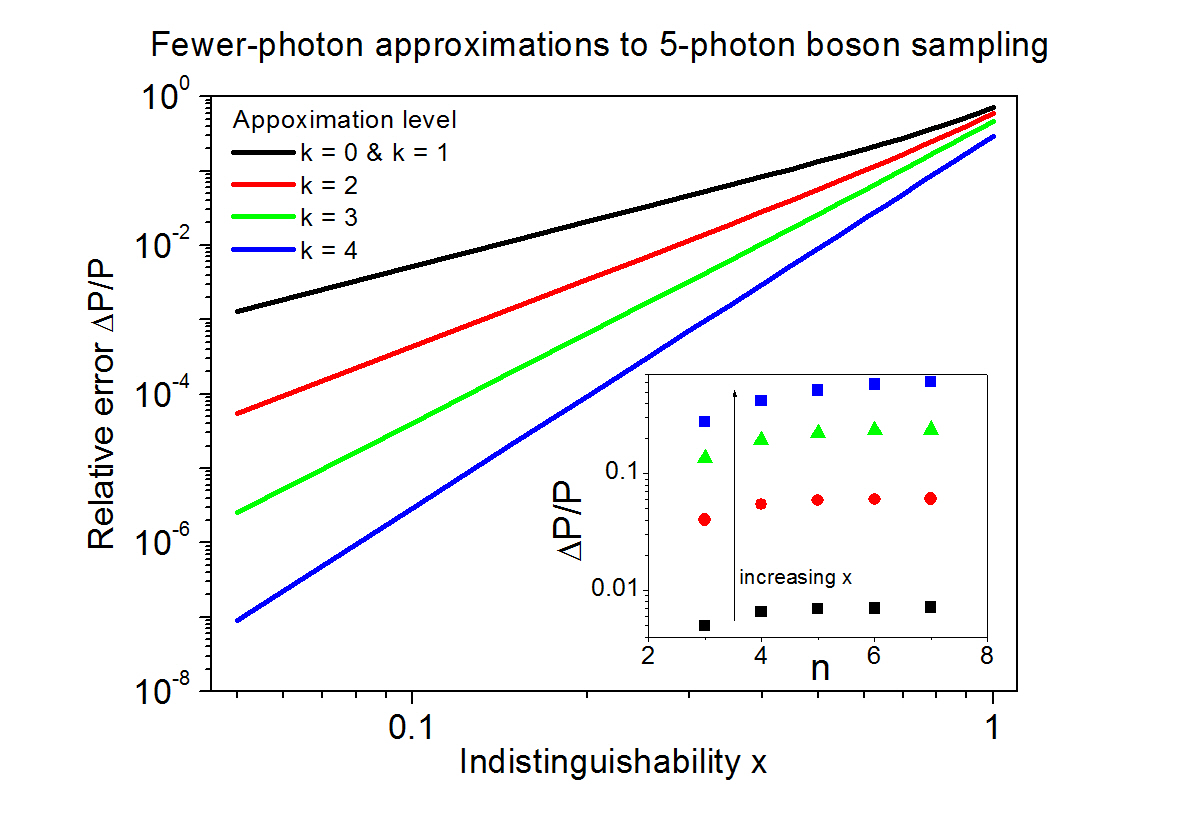}\caption{Relative error $\Delta P_{k}/P(x=0),$ where $P$ is the probability
of observing some outcome of the boson sampler, as a function of indistinguishability,
for different values of $k$. \emph{Inset} scaling of the relative
error with overall photon number for $x=0.25$, 0.5, 0.75 and 0.95
(from bottom to top).}
\end{figure}

In Figure 2, we numerically investigate the quality of this approximation
for a 5-photon boson sampling experiment. We simulate 10.000 Haar-random
unitary matrixes of size $N=100$, from which we take the first $n$
modes as input and output without loss of generality, and computed
$P_{k}$ for values of $k$ from 1 to 4. Note that the case $k=1$
only encompasses the identity permutation ($j=0)$, which corresponds
to sampling with distinguishable photons. We plot the relative error
of our approximation, defined as $\Delta P_{k}/P_{0},$ where $\Delta P=|P_{k}-P|$
and $P_{0}=n!/N^{n}$ is the characteristic scaling in the number
of photons. For photons with indistinguishability $x=0.9$, which
corresponds roughly to a recent demonstration of five-photon boson
sampling from quantum dots \cite{Wang2017}, the approximation $k=4$
already gives an approximation with an error of 10\%. 

The inset shows how the relative error scales with the photon number
$n$. We plot $\Delta P_{2}$ as a function of the photon number $n$.
We find that the relative error saturates at moderate values of $n.$
This numerical result suggests that in the limit of large photon numbers,
the accuracy of our approximation does not depend on the number of
photons, i.e. $\Delta P_{k}/P_{0}$ is a function of $k$ and $x,$
but not of $n$ if $n$ is sufficiently large. This means that for
large enough $n,$ if our approximation $P_{k}$ satisfies a given
error threshold for a particular value of $k$, that approximation
will also work for larger photon numbers. Since the size of the complex-valued
permanents is given by $k,$ this suggests that adding more photons
does not significantly add to the cost of simulating the problem.

We now proceed to prove that this is indeed the case. To do so this,
we show two things. First, we need to show that our approximation
is efficient, in the sense that adding more photons only induces a
polynomial increase in the number of computations required. Second,
we need to show that the intuition we obtained from the numerics above
is correct, and that the relative error on our approximation does
not increase when $n$ is increased. 

We begin with the first task: counting the number of terms in eq.
3. The number of terms in the middle of the three sums of eq. 3 is
given by the the rencontres number $R_{n,n-j}=\left(\begin{array}{c}
n\\
j
\end{array}\right)!(j),$ where $!j=\lfloor j!/e\rceil$ is the subfactorial, which counts
the number of permutations which leave $n-j$ elements invariant and
scales as $n^{j}$. For each term in the middle sum, we have to evaluate
$\left(\begin{array}{c}
n\\
j
\end{array}\right)$permanents of size $j,$ in the inner sum, which is a quantity which
also grows polynomially in $n.$ The total number of complex-valued
permanents which we are required to compute for particular value of
$j$ in the outer sum is therefore $\left(\begin{array}{c}
n\\
j
\end{array}\right)^{2}!(j).$ The problem therefore scales as $n^{2j}$, and when we truncate the
terms at $k,$ the number of terms is therefore of the order of $n^{2k}$.
Using Ryser's algorithm \cite{Ryser1963} to evaluate each complex-valued
permanent takes $2^{k}k$ steps, and the whole algorithm scales as
$n^{2k}2^{k}k$, which scales polynomially in $n$ as required, provided
the choice of $k$ needed to satisfy some error bound does not depend
on $n.$

Therefore, we now consider the accuracy of the scheme. We write $P$
as a polynomial in $x:$ $P(x)=\sum c_{j}x^{j},$ where the coefficients
are given by eq. 3. We can therefore estimate the error $\Delta P_{k}/P_{0}$
by computing the error term $\sum_{j=k+1}^{n}c_{j}.$ We find that
the increase in the number of terms with $j$ is precisely balanced
out by the decrease in the magnitude of each term (see supplemental
material), and that these coefficients are given by:
\[
|c_{j}|\approx\sqrt{\left(\sum_{k=0}^{n-j}(1/k!)\right)R_{n,n-j}\left(\begin{array}{c}
n\\
j
\end{array}\right)j!}(n-j)!/2N^{n}\approx n!/2N,
\]
where in the final step we have taken the limit of large $n.$ It
should be noted that this latter expression does not depend on $j,$
and differs by a factor $1/2$ from the expectation value of fully
distinguishable boson sampling. As an illustration, Figure 3 shows
estimates of the absolute values of the coefficients $|c|_{j}$, for
$n=8.$ These are compared against a numerical simulation on 500 submatrices
of Haar-random matrices. The precise behaviour of this function is
discussed in the supplemental material. Already for $n=8,$ some points
reach the value obtained in the limit of large $n$. 

Given this result, we can write $\Delta P_{k}/P$ as a geometric series:
$\Delta P_{k}/P_{0}=\sqrt{\sum_{j=k+1}^{n}(|c_{j}|x^{j})^{2}}$. If
we take the limit $n\rightarrow\infty$, this has the finite value: 

\begin{equation}
\Delta P_{k}/P_{0}=x^{k+1}/2(1-x^{2})^{1/2}.
\end{equation}

Our algorithm for simulating boson sampling with partially distinguishable
photons is now as follows: given the desired accuracy of the simulation
and the level of indistinguishability with which the expeirment is
performed, use eq. 4 to evaluate the required value of $k.$ Next,
compute eq. 2 up to the $k$-th term, and feed the computed value
of $P_{k}$ into a classical sampling algorithm, such as the Metropolis-Hastings
algorithm \cite{Hastings1970}. Such algorithms can generate a sample
from a probability distribution, even if the number of possible outcomes
is large \cite{MartinoMetropolis}, which means that the number of
modes does not enter into the hardness of boson sampling in general
\cite{Neville2017}. 

\begin{figure}
\includegraphics[width=8.8cm]{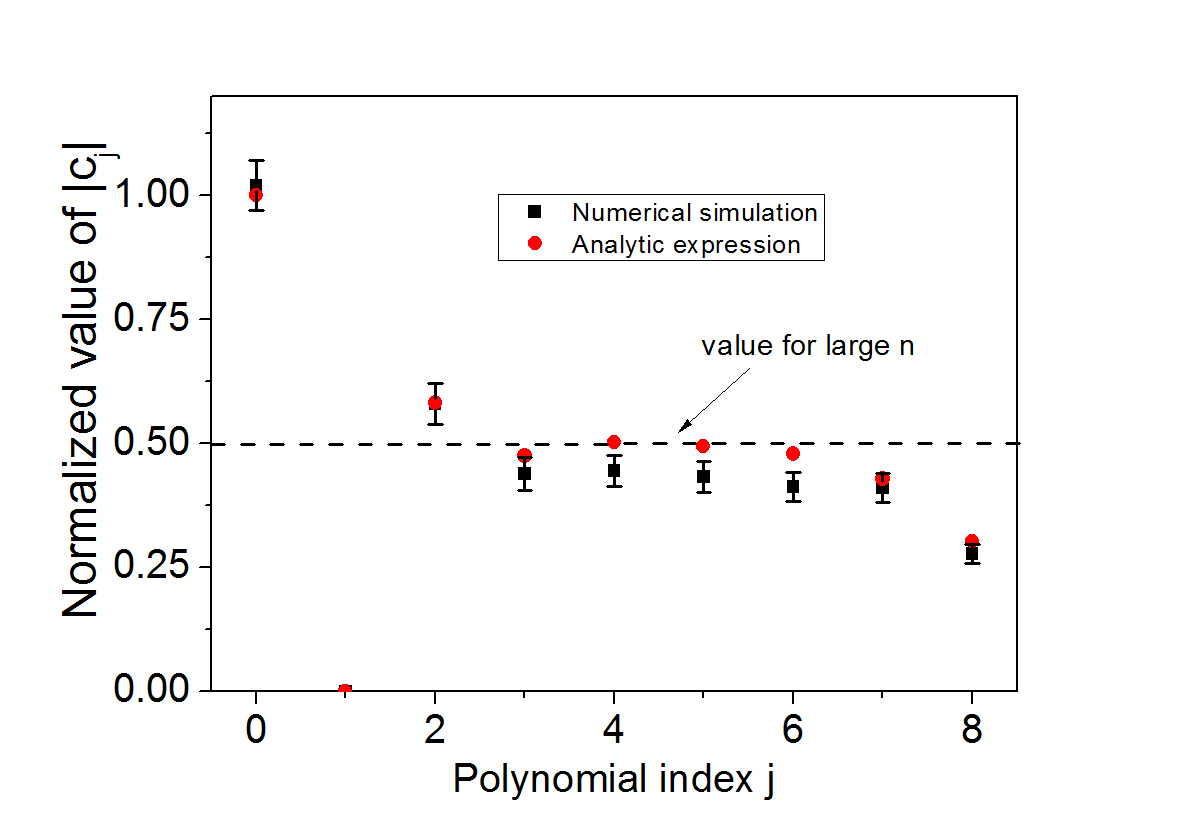}

\caption{Coefficients of the polynomial $P=\sum_{j=0}^{n}c_{j}x^{j},$ for
$n=8.$ The black squares correspond to a numerical simulation of
500 random unitaries. The red circles correspond to the prediction
from eq 4.}
\end{figure}

Finally, we detemine the regions of the parameter space where our
algorithm is efficient. The solid lines in Figure 4 show the values
of $n$ and $x$ for which eq. 4 has solutions for any $k<n$. These
lines define a threshold which must be exceeded for the complexity
of the problem not to scale polynomially with the total number of
photons. State-of-the-art supercomputers can compute permanents of
size 50 in approximately an hour \cite{Wusupercomputer}, which might
therefore be taken as an estimate of the number of photons required
to obtain a quantum advantage. Using eq. 4, we find that we require
$x=0.870$, $x=0.908$ and $x=0.947$ for $P_{49}$ to be accurate
to within 0.1\%, 1\% or 10\%, respectively. Note that if we desire
higher accuracy, our approximation fails at lower values of $x$.
We stress again that this result is a lower bound: achieving these
numbers in an experiment is no guarantee that the experiment is not
classically simulable through some more advanced algorithm. 

The dashed lines in Figure 4 show below which $n$-photon interference
can be expressed as $n-1$ photon interference, to within the given
accuracy (i.e. taking only the first term in the geometric series
for $\Delta P/P$). The area in between the dashed and solid lines
is the region of parameter space where our approximation works for
a finite value of $n$, but where the approximation will eventually
fail at large enough $n.$ 

Finally, we consider the prefactor of our algorithm. The original
expression from \cite{Tichypartial} requires the computation of roughly
$2^{n}$ permanents of size $n$, whereas the original boson sampling
proposal requires the computation of only one such permanent. If we
limit the number of computations steps we allow ourselves for computing
$P_{k}$ to be equal to the number of steps required to compute an
$n$-by-$n$ permanent, we arrive at the dash-dotted curves in Figure
4. The area in between the solid and dash-dotted curves represents
the region of the parameter space where the computational cost scales
polynomially in the number of photons, but where our approximation
might be impractical for small values of $n.$ Since Tichy's expression
requires the computation of many very similar permanents, we expect
that there is significant scope for improvement on the classical algorithm
at this point, which we leave as an open problem. 

\begin{figure}
\includegraphics{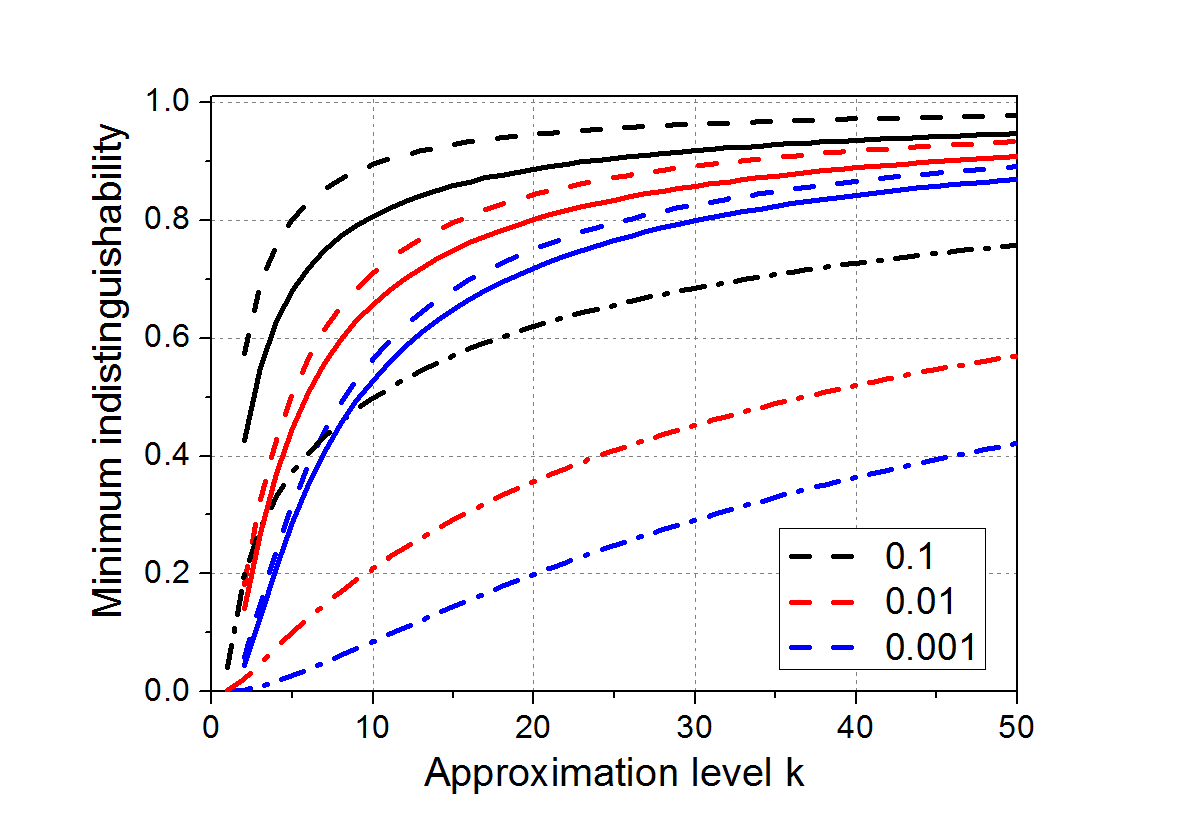}

\caption{Lower bound on the quality of photons required to achieve exponential
scaling of the hardness, as a function of the level of approxmation.
The solid lines indicate solutions of eq.4 for and demarcate the region
of polynomial scaling in $n.$ The dashed lines indicate the value
of $x$ below which $n$-photon interference can be expressed as interference
of $n-1$ photons, The dash-dotted lines indicate the values of indistinguishability
where $n$-photon interference can be so described using fewer resources
than required for the computation of an $n$-by-$n$ complex permanent. }
\end{figure}

Finally, we note that our results do not depend on the original, fairly
arbitrary parametrization of $S.$ In particular, if one has $S_{ij}=x_{ij}+\delta_{ij}(1-x_{ij}),$
with $x_{ij}<1,$ one can apply the same expansion. A similar argument
applies if the $x_{ij}$ are complex. In fact, the algorithm can be
refined to first compute those terms with large $x_{ij}$, opening
up the possibility of further approximations. Therefore our results
apply in the experimentally relevant case where all photons are not
of equal distinguishability. 

In summary, we have shown how the limited indistinguishability of
photons affects the hardness of the boson sampling problem. We have
presented a scheme that can express the probability of an outcome
in boson sampling as a sum of smaller permanents when the photons
are sufficiently distinguishable. We have demonstrated that this scheme
scales polynomially in the overall photon number, while its accuracy
does not depend on photon number. We have used this scheme to estimate
a lower bound on the indistinguishability required to achieve a quantum
advantage.
\begin{acknowledgments}
We thank Raul Garcia-Patron and Nathan Walk for useful discussions.
JJR is supported by NWO Rubicon. AM is supported by the Buckee Scholarship
from Merton College, Oxford. GT is supported by the Merton Scholarship
Fund. W.R.C. and I.A.W. acknowledge an ERC Advanced Grant (MOQUACINO).
W.S.K. is supported by EPSRC EP/M013243/1. I.A.W. acknowledges the
H2020-FETPROACT-2014 project QUCHIP (G.A. 641039).
\end{acknowledgments}


\begin{thebibliography}{10}

\bibitem{Aaronson2011}
S.~Aaronson and A.~Arkhipov,
\newblock Theory Comput. {\bf 9}, 143 (2013).

\bibitem{Spring2012}
J.~B. Spring {\em et~al.},
\newblock Science {\bf 339}, 798 (2012).

\bibitem{Broome2012}
M.~A. Broome {\em et~al.},
\newblock Science {\bf 339}, 794 (2012).

\bibitem{Crespi2013}
A.~Crespi {\em et~al.},
\newblock Nat. Photon. {\bf 7}, 545 (2013).

\bibitem{Tillmann2013}
M.~Tillmann {\em et~al.},
\newblock Nat. Photon. {\bf 7}, 540 (2013).

\bibitem{Bentivegna2015}
M.~Bentivegna {\em et~al.},
\newblock Science Advances {\bf 1}, e1400255 (2015).

\bibitem{Carolan2015}
J.~Carolan {\em et~al.},
\newblock Science {\bf 349}, 711 (2015).

\bibitem{Wang2017}
H.~Wang {\em et~al.},
\newblock Nat. Photon. {\bf 11}, 361 (2017).

\bibitem{Shchesnovich2015a}
V.~S. Shchesnovich,
\newblock Phys. Rev. A {\bf 91} (2015).

\bibitem{Arkhipov15}
A.~Arkhipov,
\newblock Phys. Rev. A {\bf 92}, 062326 (2015).

\bibitem{Aaronsonweb}
S.~Aaronson,
\newblock Scattershot bosonsampling: a new approach to scalable bosonsampling
  experiments.

\bibitem{Lund2014}
A.~P. Lund {\em et~al.},
\newblock Phys. Rev. Lett. {\bf 113}, 100502 (2014).

\bibitem{RahimiKeshari2016}
S.~Rahimi-Keshari, T.~C. Ralph, and C.~M. Caves,
\newblock Phys. Rev. X {\bf 6}, 021039 (2016).

\bibitem{Shchesnovich2014}
V.~S. Shchesnovich,
\newblock Phys. Rev. A {\bf 89} (2014).

\bibitem{Shchesnovich2015b}
V.~S. Shchesnovich,
\newblock Phys. Rev. A {\bf 91} (2015).

\bibitem{Tichypartial}
M.~C. Tichy,
\newblock Phys. Rev. A {\bf 91}, 022316 (2015).

\bibitem{Rohde2015}
P.~P. Rohde,
\newblock Phys. Rev. A {\bf 91} (2015).

\bibitem{Tamma2015}
V.~Tamma and S.~Laibacher,
\newblock Quant. Inform. Proc. {\bf 15}, 1241 (2015).

\bibitem{Tamma2015PRL}
V.~Tamma and S.~Laibacher,
\newblock Phys. Rev. Lett. {\bf 114} (2015).

\bibitem{Tamma2017}
S.~Laibacher and V.~Tamma,
\newblock (2017), arXiv:1706.05578.

\bibitem{Tillmann2015}
M.~Tillmann {\em et~al.},
\newblock Phys. Rev. X {\bf 5} (2015).

\bibitem{Jerrum2001}
M.~Jerrum, A.~Sinclair, and E.~Vigoda,
\newblock A polynomial-time approximation algorithm for the permanent of a
  matrix with non-negative entries,
\newblock in {\em Proceedings of the thirty-third annual {ACM} symposium on
  Theory of computing - {STOC} '01}, {ACM} Press, 2001.

\bibitem{Ryser1963}
H.~J. Ryser,
\newblock {\em Combinatorial Mathematics} (Mathematical Association of America,
  1963).

\bibitem{Hastings1970}
W.~K. Hastings,
\newblock Biometrika {\bf 57}, 97 (1970).

\bibitem{MartinoMetropolis}
L.~Martino and V.~Elvira,
\newblock (2017), arXiv:1704.04629.

\bibitem{Neville2017}
A.~Neville {\em et~al.},
\newblock No imminent quantum supremacy by boson sampling, 2017,
  arXiv:1705.00686.

\bibitem{Wusupercomputer}
J.~Wu {\em et~al.}, 2016, arXiv:1606.05836.

\end{thebibliography}
\end{document}